\documentstyle[12pt]{article}

\include{feynman}
\bigphotons

\newcommand{\hg}{\hat{g}}
\newcommand{\hhg}{\hat{\hat{g}}}
\newcommand{\tg}{\tilde{g}}

\renewcommand{\O}[1]{{\cal L}_{#1}}
\newcommand{\tr}{\,\rm tr\,}
\newcommand{\lag}[1]{{\cal L}_{#1}}
\newcommand{\be}{\begin{equation}}
\newcommand{\ee}{\end{equation}}
\newcommand{\bea}{\begin{eqnarray}}
\newcommand{\eea}{\end{eqnarray}}
\newcommand{\ba}[1]{\begin{array}{#1}}
\newcommand{\ea}{\end{array}}

\newcommand{\nl}{\nonumber\\}
\newcommand{\suu}{$\rm SU(2)_L \times U(1)_Y$}

\newcommand{\ep}[1]{\epsilon _{#1}}
\newcommand{\ewp}{\epsilon _{W\Phi}}
\newcommand{\ebp}{\epsilon _{B\Phi}}
\newcommand{\ew}{\epsilon _W}
\newcommand{\eeww}{$e^+ e^- \rightarrow W^+ W^-$}
\newcommand{\D}{\displaystyle}

\newcommand{\eref}[1]{(\ref{#1})}
\newcommand{\sw}[1]{\sin ^{#1}\theta _W}
\newcommand{\cw}[1]{\cos ^{#1}\theta _W}
\newcommand{\tw}[1]{\tan ^{#1}\theta _W}
\newcommand{\swc}[1]{s_W ^{#1}}
\newcommand{\cwc}[1]{c_W ^{#1}}
\newcommand{\twc}[1]{t_W ^{#1}}

\newcommand{\tfrac}[2]{\frac{\textstyle #1}{\textstyle #2}}
\newcommand{\vvvv}{V_1 V_2\rightarrow V_3 V_4}

\newcommand{\kg}{\kappa_\gamma}

\newcommand{\kz}{\kappa_Z}

\newcommand{\al}{\alpha}
\newcommand{\bt}{\beta}
\newcommand{\ga}{\gamma}
\newcommand{\de}{\delta}

\newenvironment{tabeqn}[2]{\be\renewcommand{\arraystretch}{#2}
 \arraycolsep0.5mm\begin{array}{#1}}{\end{array}\ee}
\newcommand{\bte}{\begin{tabeqn}}
\newcommand{\ete}{\end{tabeqn}}

\newcommand{\pls}{\;\;\:} 
\newcommand{\spc}{\:\,} 
\newcommand{\la}[1]{\lambda_{#1}}
 \newcommand{\vth}{\vartheta}

\newcommand{\M}[4]{ {\cal M}(\lambda_{#1}\lambda_{#2}\lambda_{#3}
\lambda_{#4})}
\newcommand{\Ms}[4]{ {\cal M}({\pls#1}{\pls#2}{\pls#3}{\pls#4}) }
\newcommand{\Mm}[4]{ {\cal M}({-\lambda_{#1}}{-\lambda_{#2}}
 {-\lambda_{#3}}{-\lambda_{#4}})}
\newcommand{\Mms}[4]{ {\cal M}({-#1}{-#2}{-#3}{-#4}) }
\newcommand{\Ml}[4]{ {\cal M}_{#1#2#3#4} }

\newcommand{\sti}{\tilde{s}}
\newcommand{\tti}{\tilde{t}}
\newcommand{\uti}{\tilde{u}}
\newcommand{\sg}{s_g}
\newcommand{\se}{s_e}
\newcommand{\sz}{s_Z}

\newenvironment{stretchfigure}[2]
 {\small\normalsize
  \begin{figure}[#2]}
 {\end{figure} \small
  \normalsize}
\newenvironment{stretchtable}[2]
 {\small\normalsize
  \begin{table}[#2]}
 {\end{table} \small
  \normalsize}

\newlength{\zeichen}
\newlength{\strich}
\newcommand{\phrd}[1]{Phys.\ Rev.\ {\bf D#1}}

\newcommand{\nphb}[1]{Nucl.\ Phys.\ {\bf B#1}}
\newcommand{\phlb}[1]{Phys.\ Lett.\ {\bf B#1}}
\newcommand{\zphc}[1]{Z.\ Phys.\ {\bf C#1}}

\small\normalsize

\newlength{\aivwidth}   
\newlength{\tmpwidth}   

\title{High-Energy Vector-Boson Scattering\\
with Non-Standard Interactions\\
and the Role of a Scalar Sector}
\author{Ingolf Kuss\thanks{Internet:kuss@physw.uni-bielefeld.de}
 \\[0.5cm]
Universit\"at Bielefeld\\Fakult\"at f\"ur Physik\\
33501 Bielefeld\\Germany}
\date{BI-TP 93/57 (condensed version)\\hep-ph/9402258\\February 1994}

\begin{document}

\maketitle
\begin{abstract}
The high-energy behavior of vector-boson scattering amplitudes is
examined within an
effective theory for non-standard self-interactions of
electroweak vector-bosons. Irrespectively of whether this theory is
brought into a gauge invariant form by including
non-standard interactions of a Higgs particle
I find that terms that grow particularly strongly with
increasing scattering energy are absent.
Different theories are compared concerning
their high-energy behavior and the appearance of divergences at the
one-loop level.
\end{abstract}
\section{Introduction}
\setcounter{figure}{0}
The standard electroweak theory \cite{sm} has been the most promising
candidate to describe the electroweak interactions
ever since it had been proven
that this theory is renormalizable \cite{thooft} and
that tree-level scattering amplitudes do not exceed the unitarity
bounds
at high energies if the mass of the Higgs boson is not too large
\cite{llewellyn,joglekar,clt}. The inclusion of a scalar
(Higgs) sector
and the generation of the vector boson masses by the spontaneous
breakdown of an underlying linearly realized local \suu\ symmetry
appears to be a necessary ingredient for a unitary and
renormalizable theory in which massive
vector particles interact with fermions. \par
In fact, the standard theory
consistently describes all currently known experimental data
and the agreement between theory and experiment is particularly
remarkable for the recent LEP~1 precision data, as
these measurements test the theory at the level of one-loop radiative
corrections.\par
However, the scalar sector of the theory and the interactions of
vector-bosons with one another have been accessed {\em only} via their
indirect, i.e., loop-induced effects on the current
observables. In view of near-future measurements of the process
\eeww\ at LEP~2 which
will be sensitive to the vector-boson self-couplings at the level of
the
Born approximation, various models that can incorporate non-standard
vector-boson self-interactions but coincide with the standard model in
the well-established sector of the interactions of the vector bosons
with the leptons and quarks have been recently under consideration.
It is not clear whether such a theory is also, like in the case
with standard self-interactions, theoretically
favored if it incorporates a scalar sector and is gauge
invariant\footnote{In this paper, the term ``gauge invariant''
means ``exhibiting a
linearly realized local \suu\ symmetry''.}.\par
Restricting ourselves to theories with the same particle content as of
the standard model, we are dealing with non-renormalizable effective
theories that have to be regularized by some ultra-violet cut-off
$\Lambda$. At the tree-level these models give rise to some
four-point amplitude that
will at high energies eventually exceed the unitarity bound
\cite{velt}. Above some scale, which is usually taken to be the
cut-off
$\Lambda$, these theories have to be embedded into some higher
theory which would
again be renormalizable and unitary.
Different approaches for constructing such effective theories exist:
\begin{enumerate}
\item Assuming that non-standard physics already appears not far above
the weak scale the scalar sector may be omitted from the theory. These
theories exhibit either {\em no} \suu\ symmetry
\cite{kmss}-\cite{bkrs}
or this symmetry is realized in a non-linear way
\cite{chiral,cvekoe}. The two cases are equivalent which can be seen
by applying a Stueckelberg transformation \cite{gkko}.\label{nosym}
\item The standard theory is adopted as the correct theory for
electroweak
interactions up to a certain scale $\Lambda$ at which new
degrees of freedom occur
and which is large compared to the mass of a Higgs particle.
The non-standard interactions may manifest themselves at energies not
too far above the weak scale as small
deviations from the standard interactions. The Lagrangian for such a
theory is an expansion in powers of $\frac{1}{\Lambda}$ around the
standard theory with gauge invariant additional operators
\cite{buwyllr}. Non-standard vector-boson self-interactions
have been recently discussed in such theories in
\cite{quad}-\cite{anom}.
\label{gaugi}
\end{enumerate}
An example of the first approach is the KMSS model
\cite{kmss} in which trilinear and quadrilinear vector-boson
self-interactions are parametrized
under a minimal set of symmetry assumptions. A two-parameter
reduction of
the KMSS model can be embedded\footnote{If not explicitly
stated otherwise, statements put up in this
paper are only valid as far as terms linear in the deviations
{from} the standard model are concerned.} \cite{anom}
into a gauge-invariant framework.
In this way one obtains an example of the second approach.
The addition of a dimension-six
single-parameter quadrupole interaction (which contains no scalar
particles and is gauge invariant in itself) to this model
leads to a general model
for vector-boson self-interactions that can be obtained by adding
gauge invariant dimension-six terms to the standard Lagrangian,
provided some reasonable
additional assumptions are made \cite{anom}. I call this model the
GINDIS (\underline{g}auge \underline{in}variant \underline{di}mension
\underline{s}ix) model.
\par
Here, I examine the behavior of vector-boson scattering-amplitudes for
large values of the center-of-mass scattering energy $s$ in the GINDIS
model. My calculation evidences that the amplitudes show
a particularly
mild growth with increasing $s$.
I further find that this behavior
is only due to the form of the vector-boson self-couplings
while the scalar sector plays no role.\par
One obvious requirement that a theory beyond the standard model has to
fulfill is that it agrees {\em at the one-loop level} with the present
precision data. If a particular model shows a dependence of
one-loop radiative corrections
on large positive powers of the cut-off,
then it can presumably not incorporate large deviations from the
standard model since these would bring the model
(in the absence of
cancellations) in conflict with present data.

\section{The Models\label{models}}
The KMSS model \cite{kmss}
describes non-standard vector-boson
self-couplings by four free parameters.
It has been derived assuming that a global SU(2) weak isospin
symmetry is broken only by electromagnetic interactions.
In particular, the symmetry is broken by a term that causes mixing
between the neutral vector-bosons.
In addition,
only dimensionless couplings and interactions that
are $P$- and $C$-even have been incorporated.
The trilinear self-couplings are described by two parameters,
$\kg$ and
$\hg$. Two other parameters, $\hhg$ and $\tg$, describe
the quadrilinear
interactions. A two parameter reduction of the KMSS model,
which I will
henceforth simply call "the KMSS model", is obtained by
imposing the conditions \cite{anom}
 \bea \hhg&=&\hg^2 , \nl
    \tg&=&0 \label{hat}.\eea
The Lagrangian of the KMSS model is given by
\bea \lag{KMSS}&=&
-ieA_\mu(W^{-\mu\nu} W^+_\nu-W^{+\mu\nu}W^-_\nu)-ie\kg
A_{\mu\nu}W^{+\mu}W^{-\nu}
\nl&&+i\left(e\frac{\sw{}}{\cw{}}-\frac{\hg}{\cw{}}\right)
Z_\mu(W^{-\mu\nu} W^+_\nu-W^{+\mu\nu}W^-_\nu)\nl&&
+i\left(e\kg\frac{\sw{}}{\cw{}}-\frac{\hg}{\cw{}}\right)
Z_{\mu\nu}W^{+\mu}W^{-\nu}
\nl&&
-e^2(A_\mu A^\mu W^+_\nu W^{-\nu} -A_\mu A_\nu W^{+\mu}
W^{-\nu})\nl&&
+2e\left(e\frac{\sw{}}{\cw{}}-\frac{\hg}{\cw{}}\right)
(A_\mu Z^\mu W^+_\nu W^{-\nu}-
\frac{1}{2}A_\mu Z_\nu (W^{+\mu} W^{-\nu}+W^{-\mu} W^{+\nu}))
\nl&&-\tfrac{\left(\hg-e\sw{}\right)^2}{\cw{2}}
(Z_\mu Z^\mu W^+_\nu W^{-\nu} -Z_\mu Z_\nu W^{+\mu} W^{-\nu})\nl&&
+\frac{1}{2}\hg^2
(W^+_\mu W^-_\nu W^{+\mu}W^{-\nu}-W^+_\mu W^-_\nu W^{-\mu}W^{+\nu}).
\label{kmss}\eea
In \eref{kmss}, $W^+_{\mu\nu}=\partial_\mu W_\nu^+-\partial_\nu
W_\mu^+$
etc., $e$ is the electron charge and $\theta_W$ is the
weak mixing angle.\par
Non-standard Higgs interactions can be added to the KMSS Lagrangian
in such a way that a
gauge invariant model results \cite{anom}. Adding to this model a
dimension-six quadrupole interaction term $\O{W}$, introduced in
\cite{quad}, which is gauge
invariant itself, one obtains the GINDIS model which has been
thoroughly discussed in \cite{anom}. The Lagrangian of the
GINDIS model is given by
\be \lag{GINDIS}=\lag{SM}
+\ep{W\Phi}\frac{g}{M_W^2}\O{W\Phi}
+\ep{B\Phi}\frac{g'}{M_W^2}\O{B\Phi}
+\ep{W}\frac{g}{M_W^2}\O{W},
\label{lageff}\ee
where $\lag{SM}$ is the standard Lagrangian.
The Lagrangian \eref{lageff} contains the three gauge invariant
dimension-six terms
\be\O{W\Phi}=i\tr[(D_\mu\Phi)^\dagger W^{\mu\nu}(D_\nu\Phi)]
\label{owp},\ee
\be\O{B\Phi}=-\frac{1}{2}i\tr[\tau_3
(D_\mu\Phi)^\dagger(D_\nu\Phi)] B^{\mu\nu}\label{obp}\ee
and
\be \O{W}=-\frac{2}{3}i
\tr (W_\mu^{\:\nu}W_\nu^{\:\lambda}W_\lambda^{\:\mu}).
\label{ow}\ee
Here, $\Phi$ denotes the standard complex scalar Higgs doublet field,
\[ \Phi=\tfrac{1}{\sqrt{2}}\left((v+H){\bf 1}+i\phi_i\tau_i\right),\]
where $H$ is a physical Higgs field, $\frac{v}{\sqrt{2}}{\bf 1}$
is the
vacuum expectation value of the Higgs doublet, the $\phi_i$ are the
would-be Goldstone bosons, ${\bf 1}$ is the unity matrix in two
dimensions and the $\tau_i$ are the Pauli matrices.
The covariant derivative of $\Phi$ is given by
\be D_\mu \Phi=\partial_\mu\Phi+igW_\mu\Phi-\frac{i}{2}g'
\Phi\tau_3B_\mu,\ee
where
\be W_\mu=\frac{1}{2}W_{\mu i}\tau_i,\ee
denotes the non-Abelian vector field, and
\bea W_{\mu\nu}&=&\partial_\mu W_\nu-\partial_\nu W_\mu+
ig[W_\mu,W_\nu],\nl
B_{\mu\nu}&=&\partial_\mu B_\nu-\partial_\nu B_\mu\label{not}\eea
are the field strength tensors.
As usual, $g$ denotes the $\rm SU(2)_L$- and $g'$ the $\rm U(1)_Y$-coupling,
$g'=\frac{e}{\cw{}}$, $M_W$ is the mass of the charged
vector-bosons
and $B_\mu$ is the $\rm U(1)_Y$ gauge field. The traces are taken over
the $2\times 2$-matrices. The non-standard couplings are described by
the parameters $\ewp, \ebp$ and $\ew$.
In this paper, I call these parameters collectively
``the $\ep{i}$''.\par
The vector-boson self-couplings of the GINDIS model with $\ew=0$
coincide with the ones of the KMSS model, \eref{kmss}, with the
identifications
\bea \hg&=&\tfrac{e}{\sw{}}\left(1+\ewp\right)\nl
\kg&=&1+\ewp+\ebp.\label{equivcubic}\eea
The GINDIS model is the general model for vector-boson
self-interactions
that can be constructed from gauge invariant dimension-six terms,
provided interactions that violate the $C$-, $P$- or
$CP$-symmetry or contain higher
derivatives are not considered. The phenomenologically
relevant parts of the GINDIS
Lagrangian in terms of the physical fields have been given in
\cite{anom}.

\section{Vector Boson Scattering\label{kinema}}
I examine the tree-level scattering amplitudes
for all processes involving only {\em massive}\ vector bosons:\par
\medskip
\renewcommand{\arraystretch}{1.3}
\be\begin{array}{cccccccc}
W^- W^+ &\rightarrow & W^- W^+&&&&&\\
W^+ W^+ &\rightarrow & W^+ W^+ &(& W^- W^- &\rightarrow & W^- W^-&)\\
W^- W^+ &\rightarrow & Z\,Z&( &Z\,Z &\rightarrow & W^- W^+&)\\
W^- Z &\rightarrow & W^- Z&( &W^+ Z &\rightarrow & W^+ Z&)\\
Z\,Z &\rightarrow & Z\,Z&,&&&&
\label{massproc}\end{array}\ee\par\medskip
as well as for the following processes involving photons\par\medskip
\be\begin{array}{cccccccc}
\gamma\,\gamma &\rightarrow & W^- W^+ &(
&W^- W^+ &\rightarrow &\gamma\,\gamma&)\\
\gamma W^- &\rightarrow &\gamma W^- &(&\gamma W^+ &\rightarrow
&\gamma W^+&)\\
\gamma\,Z&\rightarrow &W^- W^+ &(&W^- W^+&\rightarrow &\gamma\,Z&)
.\label{photproc}
\end{array}\ee\par\medskip
The amplitudes for the processes appearing in parentheses are
related to
the corresponding process to the left of them
as discussed in Appendix \ref{sym}.\par

The relevant vertices following from the GINDIS Lagrangian
\eref{lageff}
are given in
Appendix \ref{vert}.
The Feynman diagrams can be classified according to the following
scheme:
\begin{itemize}
\item Diagrams with a virtual vector boson ( denoted by {\bf V} in
 Figure \ref{fdwwww})
\item Diagrams with a four-boson vertex ( denoted by {\bf C} )
\item Diagrams with a virtual Higgs boson ( denoted by {\bf H} )
\end{itemize}
Since the calculations are performed in the unitary gauge,
there are no
diagrams with would-be-Goldstone bosons.
For the process $W^- W^+\to W^- W^+$, the diagrams are shown in Figure
\ref{fdwwww}.
There and in the general process $\vvvv$, $p_i$ is the
four-momentum of particle
$V_i$ $(i=1,2,3,4)$ and I use the Mandelstam variables:
\be s\equiv(p_1+p_2)^2,\quad
t\equiv(p_1-p_3)^2,\quad
u\equiv(p_1-p_4)^2.\label{mandelstam}\ee\par
In Figure \ref{fdwwww},
incoming particles are to the left and outgoing particles
to the right.

\begin{stretchfigure}{1}{h}
\unitlength 0.01pt
\begin{picture}(40000,48000)(-5000,0)
\begin{footnotesize}
 \put(0,33250){\begin{picture}(20000,12500)
  \put(0,11500){\large\bf V}
  \drawline\photon[\SE\REG](3000,9500)[6]
  \put(\pbackx,\pbacky){\circle*{300}}
  \global\advance\pfrontx by -1400
  \global\advance\pfronty by 300
  \put(\pfrontx,\pfronty){${\rm W^-}(p_1)$}
  \drawline\photon[\SW\REG](\photonbackx,\photonbacky)[6]
  \global\advance\pbackx by -1400
  \global\advance\pbacky by -1000
  \put(\pbackx,\pbacky){${\rm W^+}(p_2)$}
  \drawline\photon[\E\REG](\pfrontx,\pfronty)[6]
  \put(\pbackx,\pbacky){\circle*{300}}
  \global\advance\pmidx by -500
  \global\advance\pmidy by -1100
  \put(\pmidx,\pmidy){$\gamma,{\rm Z^0}$}
  \global\advance\pmidx by 400
  \global\advance\pmidy by 1500
  \put(\pmidx,\pmidy){s}
  \drawline\photon[\NE\REG](\pbackx,\pbacky)[6]
  \global\advance\pbackx by 100
  \global\advance\pbacky by 200
  \put(\pbackx,\pbacky){${\rm W^-}(p_3 )$}
  \drawline\photon[\SE\REG](\pfrontx,\pfronty)[6]
  \global\advance\pbackx by 100
  \global\advance\pbacky by -1000
  \put(\pbackx,\pbacky){${\rm W^+}(p_4 )$}
 \end{picture}}
 \put(20000,31000){\begin{picture}(20000,17000)
  \put(0,8000){+}
  \drawline\photon[\SE\REG](3000,14000)[6]
  \put(\pbackx,\pbacky){\circle*{300}}
  \global\advance\pfrontx by -1400
  \global\advance\pfronty by 300
  \put(\pfrontx,\pfronty){${\rm W^-}(p_1)$}
  \global\Xone=\photonbackx
  \global\Yone=\photonbacky
  \drawline\photon[\S\REG](\photonbackx,\photonbacky)[6]
  \put(\pbackx,\pbacky){\circle*{300}}
  \global\advance\pmidx by -1300
  \global\advance\pmidy by -300
  \put(\pmidx,\pmidy){t}
  \global\advance\pmidx by 2000
  \put(\pmidx,\pmidy){$\gamma, {\rm Z^0}$}
  \drawline\photon[\SW\REG](\pbackx,\pbacky)[6]
  \global\advance\pbackx by -1400
  \global\advance\pbacky by -1000
  \put(\pbackx,\pbacky){${\rm W^+}(p_2)$}
  \drawline\photon[\SE\REG](\pfrontx,\pfronty)[6]
  \global\advance\pbackx by 100
  \global\advance\pbacky by -1000
  \put(\pbackx,\pbacky){${\rm W^+}(p_4 )$}
  \drawline\photon[\NE\REG](\Xone,\Yone)[6]
  \global\advance\pbackx by 100
  \global\advance\pbacky by 200
  \put(\pbackx,\pbacky){${\rm W^-}(p_3 )$}
 \end{picture}}
 \put(0,18750){\begin{picture}(20000,12500)
  \put(0,11500){\large\bf C}
  \drawline\photon[\SE\REG](3000,9500)[6]
  \put(\pbackx,\pbacky){\circle*{300}}
  \global\advance\pfrontx by -1400
  \global\advance\pfronty by 300
  \put(\pfrontx,\pfronty){${\rm W^-}(p_1)$}
  \drawline\photon[\SW\REG](\photonbackx,\photonbacky)[6]
  \global\advance\pbackx by -1400
  \global\advance\pbacky by -1000
  \put(\pbackx,\pbacky){${\rm W^+}(p_2)$}
  \drawline\photon[\NE\REG](\pfrontx,\pfronty)[6]
  \global\advance\pbackx by 100
  \global\advance\pbacky by 200
  \put(\pbackx,\pbacky){${\rm W^-}(p_3 )$}
  \drawline\photon[\SE\REG](\pfrontx,\pfronty)[6]
  \global\advance\pbackx by 100
  \global\advance\pbacky by -1000
  \put(\pbackx,\pbacky){${\rm W^+}(p_4 )$}
 \end{picture}}
 \put(0,4250){\begin{picture}(20000,12500)
  \put(0,11500){\large\bf H}
  \drawline\photon[\SE\REG](3000,9500)[6]
  \put(\pbackx,\pbacky){\circle*{300}}
  \global\advance\pfrontx by -1400
  \global\advance\pfronty by 300
  \put(\pfrontx,\pfronty){${\rm W^-}(p_1)$}
  \drawline\photon[\SW\REG](\photonbackx,\photonbacky)[6]
  \global\advance\pbackx by -1400
  \global\advance\pbacky by -1000
  \put(\pbackx,\pbacky){${\rm W^+}(p_2)$}
  \drawline\scalar[\E\REG](\pfrontx,\pfronty)[3]
  \put(\pbackx,\pbacky){\circle*{300}}
  \global\advance\pmidx by -300
  \global\advance\pmidy by -1200
  \put(\pmidx,\pmidy){H}
  \global\advance\pmidx by 150
  \global\advance\pmidy by 1600
  \put(\pmidx,\pmidy){s}
  \drawline\photon[\NE\REG](\pbackx,\pbacky)[6]
  \global\advance\pbackx by 100
  \global\advance\pbacky by 200
  \put(\pbackx,\pbacky){${\rm W^-}(p_3 )$}
  \drawline\photon[\SE\REG](\pfrontx,\pfronty)[6]
  \global\advance\pbackx by 100
  \global\advance\pbacky by -1000
  \put(\pbackx,\pbacky){${\rm W^+}(p_4 )$}
 \end{picture}}
 \put(20000,2000){\begin{picture}(20000,17000)
  \put(0,8000){+}
  \drawline\photon[\SE\REG](3000,14000)[6]
  \put(\pbackx,\pbacky){\circle*{300}}
  \global\advance\pfrontx by -1400
  \global\advance\pfronty by 300
  \put(\pfrontx,\pfronty){${\rm W^-}(p_1)$}
  \global\Xone=\photonbackx
  \global\Yone=\photonbacky
  \drawline\scalar[\S\REG](\photonbackx,\photonbacky)[3]
  \put(\pbackx,\pbacky){\circle*{300}}
  \global\advance\pmidx by -1300
  \global\advance\pmidy by -300
  \put(\pmidx,\pmidy){t}
  \global\advance\pmidx by 2000
  \put(\pmidx,\pmidy){H}
  \drawline\photon[\SW\REG](\pbackx,\pbacky)[6]
  \global\advance\pbackx by -1400
  \global\advance\pbacky by -1000
  \put(\pbackx,\pbacky){${\rm W^+}(p_2)$}
  \drawline\photon[\SE\REG](\pfrontx,\pfronty)[6]
  \global\advance\pbackx by 100
  \global\advance\pbacky by -1000
  \put(\pbackx,\pbacky){${\rm W^+}(p_4 )$}
  \drawline\photon[\NE\REG](\Xone,\Yone)[6]
  \global\advance\pbackx by 100
  \global\advance\pbacky by 200
  \put(\pbackx,\pbacky){${\rm W^-}(p_3 )$}
 \end{picture}}
\end{footnotesize}
\end{picture}
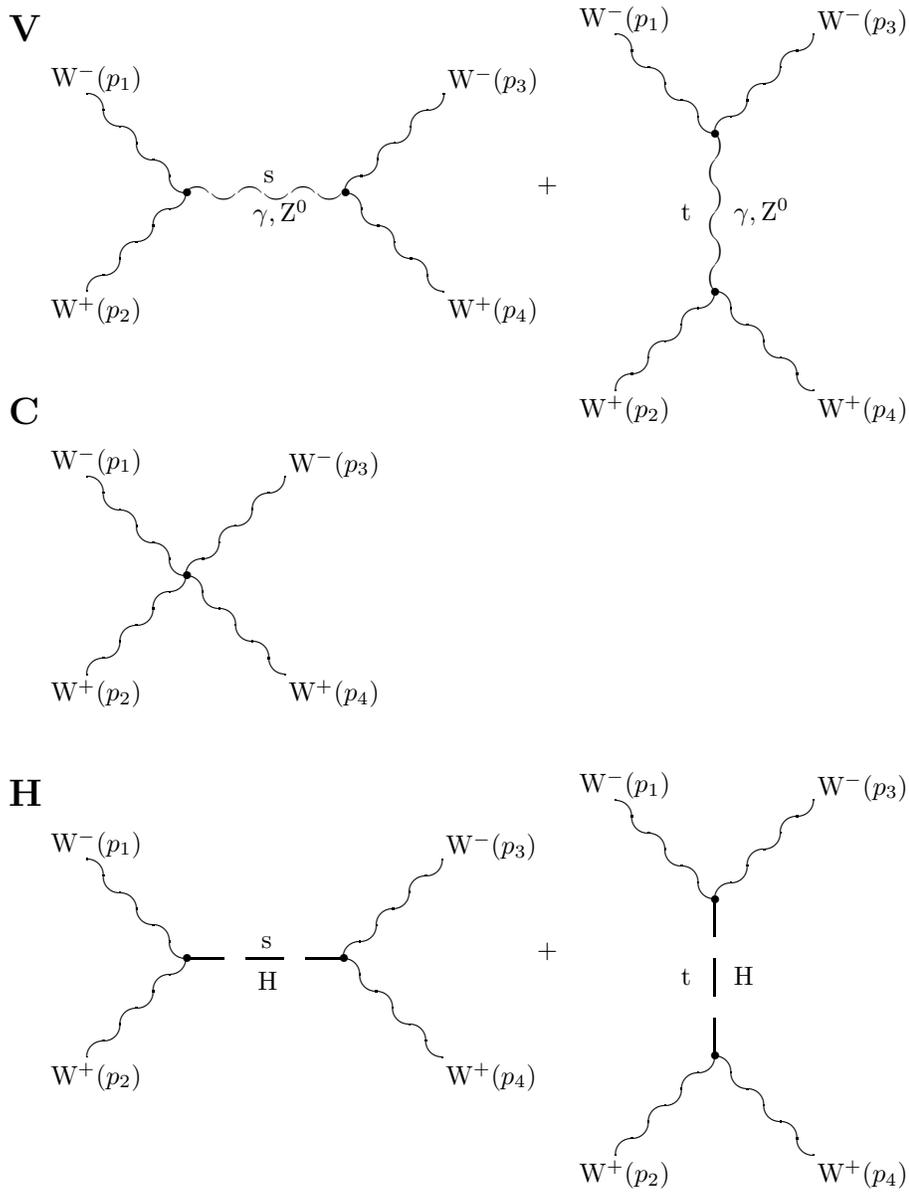
\caption[Feynman diagrams for $W^-W^+\to W^-W^+$]{
\label{fdwwww} Feynman diagrams for $W^- W^+ \rightarrow W^- W^+$ in
the tree approximation}
\end{stretchfigure}

For the general process
an amplitude ${\cal M}$ is a sum of contributions from the
three sets of graphs,
\be{\cal M}={\cal M}_V+{\cal M}_C+{\cal M}_H.\label{sets}\ee
The ${\cal M}_i$, $i=V,C$ or $H$, can in turn be written
as a product of
polarization vectors and a part which is independent of the particles'
polarizations,
\be{\cal M}_i={\cal A}^i_{\al\bt\ga\de}(p_1,p_2,p_3,p_4)
\ep{1}^\al(\lambda_1)\ep{2}^\bt(\lambda_2)\ep{3}^{\ast\ga}
(\lambda_3)\ep{4}^{\ast\de}(\lambda_4).\label{genamp}\ee
In \eref{genamp}, $\ep{j}(\lambda)$ is a polarization vector for
particle $j$ with helicity $\lambda$ and $\alpha,\gamma,\beta,\delta$
are Lorentz indices.
I use the phase conventions of Jacob and Wick
\cite{jacobwick} for the helicity eigenstates.
The appropriate polarization vectors $\ep{j}(\lambda)$
can be found in \cite{dipl}.\par
The high-energy amplitudes
are listed in Appendix \ref{amplist}.
I have also calculated the terms
bilinear in the $\ep{i}$ and find that they grow at most as
$\ep{i}\ep{j} {O}(s^2)$.\par
I turn to an investigation of the cancellations that take place among
the sets of diagrams \eref{sets} (compare \cite{joglekar}
for a similar analysis in
the standard model).
I have analyzed all amplitudes for the processes \eref{massproc} and
\eref{photproc} and
find that the sum of the vector-exchange diagrams for any amplitude
grows at most as
\[ {\cal M}_V=s^2+\ewp s^2+\ebp s+\ew s^2,\]
where $s^2$ is to be understood as ${\cal O}(s^2)$ etc.
All of these powers typically appear when all external
particles are in
the longitudinal polarization state.\par
The contact graphs get no contribution from $\ebp$, since $\O{B\Phi}$
does not contain quartic interactions. I find a growth as
\[ {\cal M}_C=s^2+\ewp s^2+\ew s^2,\]
or a more decent growth for particular amplitudes. The sum
 of vector-exchange and contact diagrams is found
to grow as
\[{\cal M}_V+{\cal M}_C=s+\ewp s+\ebp s+\ew s,\]
or more decent. The most important result can be stated here:
All the $s^2$ powers vanish already in the sum of
vector-exchange and contact diagrams. This fact holds
for all three $\ep{i}$ and in all amplitudes. If all external
particles are in the longitudinal polarization state,
there is for many amplitudes a cancellation in this
sum of one power
of $s$ in the standard term (see \cite{joglekar}).
Simultaneously there is a cancellation of one power of $s$
in the $\ewp$- and
the $\ew$-terms,
which demonstrates the special form of $\O{W\Phi}$ and $\O{W}$
as far as vector-boson self-couplings are concerned.\par
Finally, the Higgs diagrams do not depend on $\ew$ and grow at most as
\[{\cal M}_H=s+\ewp s+\ebp s,\]
and frequently they are only ${\cal O}(s^0)$.

Adding the Higgs graphs to ${\cal M}_V+ {\cal M}_C$
in order to obtain the complete amplitude, the residual
positive powers
of $s$ in the standard terms are
cancelled, so that the unitarity limit for partial wave amplitudes is
not exceeded in the standard theory at energies large compared to the
Higgs mass. This cancellation
only takes place for amplitudes in which all external
 particles are in the longitudinal
polarization state, because in the other
amplitudes the standard terms are already ${\cal O}(s^0)$.
All other effects of adding the Higgs
contribution are non-systematic: Sometimes powers of $\ewp s, \ebp
s, \ewp \sqrt{s}$ or $\ebp \sqrt{s}$ are re-introduced,
while sometimes the terms
growing as $\ebp s$ disappear.\par
One thus obtains the result,
\be{\cal M} =\ewp s+\ebp s +\ew s,\ee
or a more decent growth.
Concluding this analysis, the $\ep{i} {\cal O}(s)$ behavior is
entirely due
to the form of the non-standard vector-boson self-interactions.
The non-standard Higgs interactions yield terms of
$\ep{i} {\cal O}(s)$
but do not
change the high-energy behavior. The inclusion
of {\em a scalar sector}
in non-standard interactions is thus {\em of no relevance} as far
as the high-energy behavior of the theory is concerned.\par
I compare my result to an
analysis \cite{bks} of vector-boson scattering amplitudes
in the four-parameter
KMSS model in which the authors did \underline{not} restrict
themselves to
small values of the
$\ep{i}$. It has been shown there that
a one-parameter reduction of this model, the BKS model,
exists, in which
terms that grow with ${\cal O}(s^2)$ or stronger are absent in the
amplitudes. The Lagrangian of the BKS model can be obtained from
\eref{kmss} by eliminating $\hg$ in favor of $\kg$ by the relation
\be \hg=\tfrac{e}{\sw{}}\kg.\label{hgkg}\ee
I note that from
\eref{equivcubic} and \eref{hgkg} one sees that the BKS model is
equivalent to the GINDIS model \eref{lageff} with
\be \ewp=\kg-1\quad{\rm and}\quad\ebp=\ew=0,\ee
as far as the vector-boson self-couplings are
concerned. Thus, the model with only $\ewp$ should yield
only amplitudes
that grow at most as ${\cal O}(s)$ if the
Higgs interactions are turned off.
My calculations show that the terms $\ewp {\cal O}(s)$ remain
if the Higgs interactions are added.
I expected this latter fact from a result in \cite{bks}.
The authors of this
reference showed that in the BKS model unitarity is violated
in partial
waves with angular momentum $J > 1$ so that the addition of only a
scalar particle is not sufficient to restore unitarity.\par
Considering the equivalence of the KMSS model and the
GINDIS model without Higgs
interactions and without $\O{W}$,
one expects that
amplitudes for a model with only $\O{B\Phi}$ are in
general ${\cal M}= {\cal O}(s^2)$, since the BKS model
(which is equivalent
to taking only $\O{W\Phi}$) is the {\em only} model that
can be embedded
into the KMSS model in which the amplitudes are only ${\cal O}(s)$
even bilinear in the deviations from the standard couplings.
It is therefore worth remarking here that
$s^2$ terms appear only quadratically in $\ebp$.\par

\section{Conclusions}
It is well-known that the inclusion of the Higgs-extension in the
standard model is crucial to ensure the perturbative unitarity
as well as the renormalizability of the theory.
Concerning the dependence of one-loop radiative
corrections to four-fermion scattering amplitudes on the mass of the
Higgs boson, the standard theory only shows a mild, logarithmic
$M_H$-dependence.\par
The role of the Higgs-sector in the standard theory can be studied by
looking at the corresponding behavior in the non-linear sigma-model
\cite{appe}, which is obtained by integrating out the physical Higgs
particle of the standard theory. The non-linear sigma-model in the
unitary gauge
corresponds to the case of {\em no} Higgs particle, or,
equivalently, to
the limit of an infinite mass of the Higgs particle in the standard
model. In the non-linear sigma-model,
tree-amplitudes grow as ${\cal O}(s)$. In contrast
to the standard model, this model is non-renormalizable. It has a
logarithmic cut-off dependence. \par
I have investigated the role of the Higgs-extension in effective
theories with non-standard vector-boson self-couplings.
These theories are
non-renormalizable even when a Higgs-extension is included.
Only terms at most linear in
the deviations from the standard couplings have been considered.
This restriction is also
explicitly assumed in the following discussion, if not otherwise
specified.\par
I start with the KMSS model, \eref{kmss}.
This model is equivalent to the GINDIS model with $\ew=0$,
$\ewp\neq 0$, $\ebp\neq 0$ and no Higgs interactions. We saw
that vector-boson
scattering amplitudes grow at most as ${\cal O}(s)$.
The ${\cal O}(s)$-growth remains if the Higgs interactions are added.
Thus, in distinction
{from} the case of the models with {\em standard} vector-boson
self-couplings, the omission of the Higgs-extension in the
non-standard
interactions does {\em not}
change the high-energy behavior of tree-amplitudes.
If we add the quadrupole interaction $\O{W}$ we also find at most an
${O}(s)$-growth.\par
As to loop effects, a complete analysis of one-loop corrections to
four-fermion scattering amplitudes in a model
which reduces to the GINDIS model in a special case
has been presented in \cite{haze}. It is shown there that
the effects of the
non-standard terms can be described by cut-off dependent
(renormalized)
coefficients of other dimension-six terms which have
tree level effects and
by a renormalization of the standard model parameters. I note that
only the dependence of the coefficients on the scale
$\Lambda$ can be determined
unless one knows the underlying renormalizable theory \cite{santa}.
If the Higgs-interactions are excluded (KMSS
model), the renormalized coefficients are proportional to $\Lambda^2$
and
$\ln\Lambda$. When the Higgs sector is included, the quadratic
$\Lambda$-dependence disappears and only an $\ln\Lambda$-dependence
remains. In addition, a quadratic $M_H$-dependence appears. This
behavior is similar to the replacement rule $M_H\to\Lambda$ when going
{from} the standard model to the non-linear sigma-model,
although for the
effective theory this replacement does not quantitatively
reproduce the
heavy-Higgs limit.\par
The behavior of the different models is summarized
in Table \ref{sum}.\par
Finally, I note that in the four-parameter KMSS model,
which can not be embedded into a
gauge invariant framework without taking
dimension-eight terms, we have an
${O}(s^2)$ growth of amplitudes.
The $\rho$-parameter depends only quadratically on the cut-off
\cite{kmsscut}, but the $\Lambda$-dependence  of the other one-loop
contributions has not yet been investigated.
\begin{stretchtable}{1}{h}
\renewcommand{\arraystretch}{1.1}
\begin{minipage}{\textwidth}
\[\begin{array}{|c||c|c|c||c|c|cc|}
\hline
&\multicolumn{3}{c||}{ {\rm\bf No~Higgs} }&\multicolumn{4}{c|}{
{\rm\bf Linear~Higgs~Sector}}\\ \hline
&{\rm Model}&{\rm Amplitudes}&{\rm Loops}&{\rm Model}&
{\rm Amplitudes}&\multicolumn{2}{|c|}{\rm Loops}\\
\raisebox{1.5ex}[-1.5ex]{\rm Self-Interactions}&&&&&&M_H&
{\rm Cut{-}off}\\ \hline
{\rm\bf Standard}&{\rm a)}&{O}(s)&\ln\Lambda&{\rm b)}&{O}(s^0)&\ln
M_H&\mbox{--}\\ \hline
&{\rm c)}&{O}(s)&\Lambda^2,\ln\Lambda&{\rm d)}&{O}(s)&M_H^2
&\ln\Lambda\\
\raisebox{1.5ex}[-1.5ex]{\rm\bf Non-Standard}&{\rm
e)}&{O}(s^2)&\
&&&&\\ \hline
\end{array}\]
\protect\caption[Amplitude growth, $\Lambda$- and $M_H$-dependence]{
Growth of tree-amplitudes, cut-off- and $M_H$-dependence in various
theories.\\
a) Non-linear $\sigma$-model\\
b) Standard model\\
c) BKS model, two-parameter KMSS model\\
d) Linearly \suu\ invariant dimension-six extension of the standard
model (GINDIS)\\
e) four-parameter KMSS model\label{sum} }
\end{minipage}
\end{stretchtable}

\section*{Acknowledgements}
\addcontentsline{toc}{chapter}{Acknowledgement}
I would like to thank Dieter Schildknecht for suggesting
this work to me and for continuous advice.
I thank Mikhail Bilenky, Carsten Grosse-Knetter and
Hubert Spiesberger
for useful discussions and help.

\clearpage
\begin{appendix}
\section*{Appendices}
\renewcommand{\thefigure}{\thesection.\arabic{figure}}
\section{Vertices\label{vert}}
\setcounter{figure}{0}
The vertices needed for the computation of the amplitudes
can be classified according to
\begin{itemize}
\item Vertices involving three vector bosons (Figure \ref{vert3b})
\item Vertices involving four vector bosons (Figure \ref{vert4b})
\item Vertices involving one Higgs scalar and two vector bosons
 (Figure \ref{verthb})
\end{itemize}
In Figures \ref{vert3b} to \ref{verthb}, the vertices
are explicitely given.
All particles are understood to be entering the vertex.
Vertex functions involving outgoing particles can easily
be constructed
by replacing an incoming particle by the
outgoing antiparticle and simultaneously replacing
the particle's four-momentum by its negative four-momentum.

\unitlength 0.01pt
\begin{stretchfigure}{1}{h}
\begin{picture}(39000,32500)
 \put(0,15000){\begin{picture}(18000,12000)
  \drawline\photon[\W\REG](10000,6000)[6]
  \global\advance\pbackx by -3700
  \global\advance\pbacky by -300
  \put(\pbackx,\pbacky){${\rm V}_\alpha (k_1)$}
  \drawline\photon[\NE\REG](\pfrontx,\pfronty)[6]
  \global\advance\pbackx by 300
  \global\advance\pbacky by -300
  \put(\pbackx,\pbacky){${\rm W}^+ _\beta (k_2)$}
  \drawline\photon[\SE\REG](\pfrontx,\pfronty)[6]
  \global\advance\pbackx by 300
  \global\advance\pbacky by -300
  \put(\pbackx,\pbacky){${\rm W}^- _\gamma (k_3)$}
 \end{picture}}
 \put(18000,12500){\makebox(25000,20000){
  \renewcommand{\arraystretch}{1.5}
  \arraycolsep01pt
  $\begin{array}{rcccl}
  &\multicolumn{4}{l}{V_{V\,W\,W}(\alpha ,\beta
 ,\gamma ,k_1 ,k_2 ,k_3 ,
   g_V ,\kappa _V ,y_V )} \vspace{15pt} \\
   =&&ieg_V \Big[&-&g^{\alpha\beta}k_2^\gamma +g^{\beta\gamma}
 (k_2 -k_3
   )^\alpha +g^{\gamma\alpha}k_3^\beta\\
   &&&+&\kappa _V (g^{\alpha\beta}k_1^\gamma
 -g^{\gamma\alpha}k_1^\beta
   )\Big]\\
   &+&ie\tfrac{y_V}{M_W^2}\Big[&&k_1^\gamma
 k_2^\alpha k_3^\beta -k_1^\beta
   k_2^\gamma k_3^\alpha\\
   &&&+&k_1 \cdot k_2 (k_3^\alpha g^{\beta\gamma}-k_3^\beta
   g^{\alpha\gamma})\\
   &&&+&k_1 \cdot k_3 (k_2^\gamma g^{\alpha\beta}-k_2^\alpha
   g^{\beta\gamma})\\
   &&&+&k_2 \cdot k_3 (k_1^\beta g^{\alpha\gamma}-k_1^\gamma
   g^{\alpha\beta})\Big]
  \end{array}$}
 }
 \put(0,0){\makebox(40000,15000){
  \renewcommand{\arraystretch}{2.2}
  $\begin{array}{rrclrcl}
  {\rm with}&\multicolumn{6}{l}{V=\;\gamma\quad{\rm or}\quad Z}\\
  {\rm and}&g_\gamma&=&1,&g_Z&=&\tfrac{\cwc{2}+\ep{W\Phi}}
 {\swc{}\cwc{}}\\
  &\kappa _\gamma&=&1+\ep{W\Phi}+\ep{B\Phi},&
 \kz&=&\tfrac{\cwc{2}}{\cwc{2}
   +\ep{W\Phi}}\left(
   1+\ep{W\Phi}-\tfrac{\swc{2}}{\cwc{2}}\ep{B\Phi}\right)\\
  &y_\gamma&=&\ep{W},&y_Z&=&\tfrac{\cwc{}}{\swc{}}\ep{W}
  \end{array}$}
 }
\end{picture}
\caption{Three-Boson Vertex\label{vert3b}}
\end{stretchfigure}

\begin{stretchfigure}{1}{h}
\begin{picture}(39000,15000)
 \put(0,0){\begin{picture}(18000,12000)
  \drawline\photon[\NW\REG](10000,6000)[6]
  \global\advance\pbackx by -3500
  \global\advance\pbacky by -300
  \put(\pbackx,\pbacky){$\gamma _\alpha (k_1)$}
  \drawline\photon[\SW\REG](\pfrontx,\pfronty)[6]
  \global\advance\pbackx by -3500
  \global\advance\pbacky by -300
  \put(\pbackx,\pbacky){$\gamma _\beta (k_2)$}
  \drawline\photon[\NE\REG](\pfrontx,\pfronty)[6]
  \global\advance\pbackx by 300
  \global\advance\pbacky by -300
  \put(\pbackx,\pbacky){${\rm W}^+ _\gamma (k_3)$}
  \drawline\photon[\SE\REG](\pfrontx,\pfronty)[6]
  \global\advance\pbackx by 300
  \global\advance\pbacky by -300
  \put(\pbackx,\pbacky){${\rm W}^- _\delta (k_4)$}
 \end{picture}}
 \put(18000,0){\makebox(23000,12000){
  \renewcommand{\arraystretch}{1.5}
  \arraycolsep01pt
  $\begin{array}{rccl}
  &\multicolumn{3}{l}{V_{\gamma\,\gamma\,W\,W}(\alpha
  ,\beta ,\gamma , \delta,
    k_1 ,k_2 ,k_3 ,k_4 )} \vspace{15pt} \\
   =&\,ie^2 \Big[&&g^{\alpha\gamma}
 g^{\beta\delta}+g^{\alpha\delta}g^{\beta
     \gamma}-2g^{\alpha\beta}g^{\gamma\delta}\\
   &&+&\tfrac{\ep{W}}{M_W^2}F^
 {\alpha\beta\gamma\delta}(k_1,k_2,k_3,k_4)
   \Big]
  \end{array}$}
 }
\end{picture}

\begin{picture}(39000,15000)
 \put(0,0){\begin{picture}(18000,12000)
  \drawline\photon[\NW\REG](10000,6000)[6]
  \global\advance\pbackx by -3500
  \global\advance\pbacky by -300
  \put(\pbackx,\pbacky){${\rm Z}_\alpha (k_1)$}
  \drawline\photon[\SW\REG](\pfrontx,\pfronty)[6]
  \global\advance\pbackx by -3500
  \global\advance\pbacky by -300
  \put(\pbackx,\pbacky){$\gamma _\beta (k_2)$}
  \drawline\photon[\NE\REG](\pfrontx,\pfronty)[6]
  \global\advance\pbackx by 300
  \global\advance\pbacky by -300
  \put(\pbackx,\pbacky){${\rm W}^+ _\gamma (k_3)$}
  \drawline\photon[\SE\REG](\pfrontx,\pfronty)[6]
  \global\advance\pbackx by 300
  \global\advance\pbacky by -300
  \put(\pbackx,\pbacky){${\rm W}^- _\delta (k_4)$}
 \end{picture}}
 \put(18000,0){\makebox(23000,12000){
  \renewcommand{\arraystretch}{1.5}
  \arraycolsep01pt
  $\begin{array}{rccl}
  &\multicolumn{3}{l}{V_{Z\,\gamma\,W\,W}(\alpha
 ,\beta ,\gamma , \delta,
    k_1 ,k_2 ,k_3 ,k_4 )} \vspace{15pt} \\
   =&\,ie^2\,\tfrac{\cwc{}}{\swc{}} \Big[&&\left(
 1+\tfrac{\ep{W\Phi}}{\cwc{2}}
    \right)\\
    &&&\cdot (g^{\alpha\gamma}g^{\beta\delta}+g^{\alpha\delta}g^{\beta
     \gamma}-2g^{\alpha\beta}g^{\gamma\delta})\\
   &&+&\tfrac{\ep{W}}{M_W^2}F^
 {\alpha\beta\gamma\delta}(k_1,k_2,k_3,k_4)
   \Big]
  \end{array}$}
 }
\end{picture}

\begin{picture}(39000,15000)
 \put(0,0){\begin{picture}(18000,12000)
  \drawline\photon[\NW\REG](10000,6000)[6]
  \global\advance\pbackx by -3500
  \global\advance\pbacky by -300
  \put(\pbackx,\pbacky){${\rm Z}_\alpha (k_1)$}
  \drawline\photon[\SW\REG](\pfrontx,\pfronty)[6]
  \global\advance\pbackx by -3500
  \global\advance\pbacky by -300
  \put(\pbackx,\pbacky){${\rm Z}_\beta (k_2)$}
  \drawline\photon[\NE\REG](\pfrontx,\pfronty)[6]
  \global\advance\pbackx by 300
  \global\advance\pbacky by -300
  \put(\pbackx,\pbacky){${\rm W}^+ _\gamma (k_3)$}
  \drawline\photon[\SE\REG](\pfrontx,\pfronty)[6]
  \global\advance\pbackx by 300
  \global\advance\pbacky by -300
  \put(\pbackx,\pbacky){${\rm W}^- _\delta (k_4)$}
 \end{picture}}
 \put(18000,0){\makebox(23000,12000){
  \renewcommand{\arraystretch}{1.5}
  \arraycolsep01pt
  $\begin{array}{rccl}
  &\multicolumn{3}{l}{V_{Z\,Z\,W\,W}(\alpha ,\beta ,\gamma , \delta,
    k_1 ,k_2 ,k_3 ,k_4 )} \vspace{15pt} \\
   =&\,ie^2\,\tfrac{\cwc{2}}{\swc{2}} \Big[&&\left(
 1+2\tfrac{\ep{W\Phi}}{
    \cwc{2}}\right)\\
    &&&\cdot (g^{\alpha\gamma}g^{\beta\delta}+g^{\alpha\delta}g^{\beta
     \gamma}-2g^{\alpha\beta}g^{\gamma\delta})\\
   &&+&\tfrac{\ep{W}}{M_W^2}F^
 {\alpha\beta\gamma\delta}(k_1,k_2,k_3,k_4)
   \Big]
  \end{array}$}
 }
\end{picture}

\begin{picture}(39000,15000)
 \put(0,0){\begin{picture}(18000,12000)
  \drawline\photon[\NW\REG](10000,6000)[6]
  \global\advance\pbackx by -4000
  \global\advance\pbacky by -300
  \put(\pbackx,\pbacky){${\rm W}^+ _\alpha (k_1)$}
  \drawline\photon[\SW\REG](\pfrontx,\pfronty)[6]
  \global\advance\pbackx by -4000
  \global\advance\pbacky by -300
  \put(\pbackx,\pbacky){${\rm W}^- _\beta (k_2)$}
  \drawline\photon[\NE\REG](\pfrontx,\pfronty)[6]
  \global\advance\pbackx by 300
  \global\advance\pbacky by -300
  \put(\pbackx,\pbacky){${\rm W}^+ _\gamma (k_3)$}
  \drawline\photon[\SE\REG](\pfrontx,\pfronty)[6]
  \global\advance\pbackx by 300
  \global\advance\pbacky by -300
  \put(\pbackx,\pbacky){${\rm W}^- _\delta (k_4)$}
 \end{picture}}
 \put(18000,0){\makebox(23000,12000){
  \renewcommand{\arraystretch}{1.5}
  \arraycolsep01pt
  $\begin{array}{rccl}
  &\multicolumn{3}{l}{V_{W\,W\,W\,W}(\alpha ,\beta ,\gamma , \delta,
    k_1 ,k_2 ,k_3 ,k_4 )} \vspace{15pt} \\
   =&\,-ie^2\,\tfrac{1}{\swc{2}} \Big[&&\left( 1+2\ep{W\Phi}
    \right)\\
    &&&\cdot (g^{\alpha\gamma}g^{\beta\delta}+g^{\alpha\delta}g^{\beta
     \gamma}-2g^{\alpha\beta}g^{\gamma\delta})\\
   &&+&\tfrac{\ep{W}}{M_W^2}F^{\alpha\gamma\beta\delta}
 (k_1,k_3,k_2,k_4)
   \Big]
  \end{array}$}
 }
\end{picture}
\caption[Four-Boson Vertices (part 1)]{
Four-Boson Vertices (to be continued)\label{vert4b}}
\end{stretchfigure}

\begin{stretchfigure}{1}{h}
\addtocounter{figure}{-1}
\begin{picture}(40000,25000)
\put(0,0){\makebox(40000,25000){
 \renewcommand{\arraystretch}{1.5}
 \arraycolsep01pt
 $\begin{array}{rcl}
 \multicolumn{3}{l}{\rm where}\vspace{10pt}\\
 \multicolumn{3}{l}{F^{\alpha\beta\gamma\delta}(k_1,k_2,k_3,k_4)\; =}
 \vspace{10pt}\\
 \quad\quad &&g^{\alpha\beta}g^{\gamma\delta}\left( (k_1\cdot k_3)+
 (k_1\cdot k_4)+(k_2\cdot k_3)+(k_2\cdot k_4)\right)\\
 &-&g^{\alpha\gamma}g^{\beta\delta}\left( (k_1\cdot k_4)
+(k_2\cdot k_3)
 \right)\\
 &-&g^{\alpha\delta}g^{\beta\gamma}\left( (k_1\cdot k_3)
 +(k_2\cdot k_4)
 \right)\\
 &-&g^{\alpha\beta}(k_1^\gamma k_3^\delta +k_2^\gamma k_3^\delta
 +k_1^\delta k_4^\gamma +k_2^\delta k_4^\gamma)\\
 &-&g^{\gamma\delta}(k_1^\beta k_3^\alpha +k_2^\alpha k_3^\beta
 +k_1^\beta k_4^\alpha +k_2^\alpha k_4^\beta)\\
 &+&g^{\alpha\gamma}(k_1^\beta k_3^\delta -k_1^\delta k_3^\beta
 +k_2^\delta k_3^\beta +k_1^\delta k_4^\beta)\\
 &+&g^{\alpha\delta}(k_1^\beta k_4^\gamma -k_1^\gamma k_4^\beta
 +k_2^\gamma k_4^\beta + k_1^\gamma k_3^\beta)\\
 &+&g^{\beta\gamma}(k_1^\delta k_3^\alpha +k_2^\alpha k_3^\delta
 -k_2^\delta k_3^\alpha +k_2^\delta k_4^\alpha)\\
 &+&g^{\beta\delta}(k_1^\gamma k_4^\alpha +k_2^\alpha k_4^\gamma
 -k_2^\gamma k_4^\alpha +k_2^\gamma k_3^\alpha)
 \end{array}$}
}
\end{picture}
\caption[Four-Boson Vertices (part 2)]{Four-Boson Vertices (contd.) }
\end{stretchfigure}

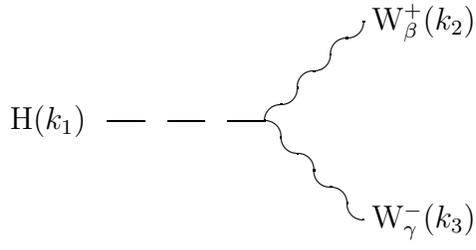
\begin{stretchfigure}{1}{h}
\begin{picture}(40000,14000)
 \put(0,0){\begin{picture}(18000,12000)
  \drawline\scalar[\W\REG](10000,6000)[3]
  \global\advance\pbackx by -3700
  \global\advance\pbacky by -300
  \put(\pbackx,\pbacky){${\rm H}(k_1)$}
  \drawline\photon[\NE\REG](\pfrontx,\pfronty)[6]
  \global\advance\pbackx by 300
  \global\advance\pbacky by -300
  \put(\pbackx,\pbacky){${\rm W}^+ _\beta (k_2)$}
  \drawline\photon[\SE\REG](\pfrontx,\pfronty)[6]
  \global\advance\pbackx by 300
  \global\advance\pbacky by -300
  \put(\pbackx,\pbacky){${\rm W}^- _\gamma (k_3)$}
 \end{picture}}
 \put(18000,0){\makebox(25000,12000){
  \renewcommand{\arraystretch}{1.5}
  \arraycolsep01pt
  $\begin{array}{rccccl}
  &\multicolumn{5}{l}{V_{H\,W\,W}(\beta ,\gamma ,k_1 ,k_2 ,k_3 ,
   )} \vspace{15pt} \\
   =&i\,g\Big[&&\multicolumn{3}{l}{M_W g^{\beta\gamma}}\\
  &&+&\tfrac{\ep{W\Phi}}{M_W}\Big( &&g^{\beta\gamma}k_1\cdot (k_2
   +k_3)\\
  &&&&-&k_2^\gamma k_1^\beta-k_3^\beta k_1^\gamma\Big) \:\Big]
  \end{array}$}
 }
\end{picture}
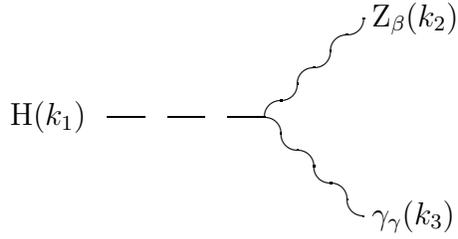
\begin{picture}(40000,14000)
 \put(0,0){\begin{picture}(18000,12000)
  \drawline\scalar[\W\REG](10000,6000)[3]
  \global\advance\pbackx by -3700
  \global\advance\pbacky by -300
  \put(\pbackx,\pbacky){${\rm H}(k_1)$}
  \drawline\photon[\NE\REG](\pfrontx,\pfronty)[6]
  \global\advance\pbackx by 300
  \global\advance\pbacky by -300
  \put(\pbackx,\pbacky){${\rm Z}_\beta (k_2)$}
  \drawline\photon[\SE\REG](\pfrontx,\pfronty)[6]
  \global\advance\pbackx by 300
  \global\advance\pbacky by -300
  \put(\pbackx,\pbacky){$\gamma _\gamma (k_3)$}
 \end{picture}}
 \put(18000,0){\makebox(25000,12000){
  \renewcommand{\arraystretch}{1.5}
  \arraycolsep01pt
  $\begin{array}{rcl}
  &\multicolumn{2}{l}{V_{H\,Z\,\gamma}(\beta ,\gamma ,k_1 ,k_2 ,k_3
   )} \vspace{15pt} \\
  =&i\,\tfrac{e}{\cwc{}}\tfrac{1}{M_W}&\left( \ep{W\Phi}-\ep{B\Phi}
  \right)\\
  &&\cdot \left( g^{\beta\gamma}(k_1\cdot k_3)
 -k_3^\beta k_1^\gamma\right)
  \end{array}$}
 }
\end{picture}
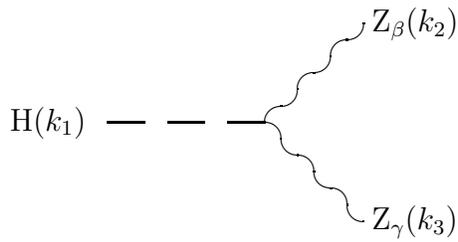
\begin{picture}(40000,14000)
 \put(0,0){\begin{picture}(18000,12000)
  \drawline\scalar[\W\REG](10000,6000)[3]
  \global\advance\pbackx by -3700
  \global\advance\pbacky by -300
  \put(\pbackx,\pbacky){${\rm H}(k_1)$}
  \drawline\photon[\NE\REG](\pfrontx,\pfronty)[6]
  \global\advance\pbackx by 300
  \global\advance\pbacky by -300
  \put(\pbackx,\pbacky){${\rm Z}_\beta (k_2)$}
  \drawline\photon[\SE\REG](\pfrontx,\pfronty)[6]
  \global\advance\pbackx by 300
  \global\advance\pbacky by -300
  \put(\pbackx,\pbacky){${\rm Z}_\gamma (k_3)$}
 \end{picture}}
 \put(18000,0){\makebox(25000,12000){
  \renewcommand{\arraystretch}{1.5}
  \arraycolsep01pt
  $\begin{array}{rcccccl}
  &\multicolumn{6}{l}{V_{H\,Z\,Z}(\beta ,\gamma ,k_1 ,k_2 ,k_3 ,
   )} \vspace{15pt} \\
   =&i\,g\Big[&&\multicolumn{4}{l}{\tfrac{M_W}{\cwc{2}}
 g^{\beta\gamma}}\\
  &&+&\tfrac{1}{M_W}&\multicolumn{3}{l}{\left(
 \ep{W\Phi}+\tfrac{\swc{2}}
  {\cwc{2}}\ep{B\Phi}\right)}\\
  &&&&\cdot \Big( &&g^{\beta\gamma}k_1\cdot (k_2
   +k_3)\\
  &&&&&-&k_2^\gamma k_1^\beta-k_3^\beta k_1^\gamma\Big) \:\Big]
  \end{array}$}
 }
\end{picture}
\caption{Vertices with One Higgs Boson and Two
 Vector Bosons\label{verthb}}
\end{stretchfigure}
\clearpage

\section[Symmetries]{Relations between Amplitudes\label{sym}}
\setcounter{figure}{0}
Given the particle types in the initial and final states, there is in
principle a number of 81 different amplitudes if all
particles are massive.
The number of distinct amplitudes
can be significantly reduced, however, if one relates
 certain amplitudes
to each other by using the fact that the $S$-matrix is invariant under
$C$-, $P$- and $T$-transformations (e.g.
 \cite{jacobwick,grs,symmetries}).
Also, amplitudes for reactions
involving different sets of particles can be related
 to each other.\par
I denote an amplitude for the reaction of the
 particles $AB\to CD$ with helicities
$\la 1,\la 2,\la 3,\la 4$ (in this order)
and center-of-mass scattering
angle $\vth$ by ${\cal M}(\la 1\la 2\la 3\la 4)(AB\to CD)(\vth)$.
In the following, the frame axes in the center-of-mass system
 are defined in such a way that the
reaction takes place in the \^{x}-\^{z} plane. Particle $A$
travels in
the positive \^{z}-direction and particle $C$ has momentum component
$p_{\rm x}\ge 0$. The scattering angle $\vth$ is
restricted to $0\le\vth\le\pi$.
In the relations I give here, I take into account the phase factors
according to the Jacob and Wick phase convention.
Derivations of the relations can be found in \cite{dipl}.\par
{}From the invariance of the $S$-matrix under
a \underline{rotation} one obtains
the relation
\be\M 1234(AB\to CD)(\vth)=(-1)^{\la1-\la2+\la4-\la3}
\M 2143(BA\to DC)(\vth)
.\ee  This relation can also be obtained from exchanging the two
particles in the initial states as well as the two particles in the
final state.\par
A \underline{parity} transformation $P$ changes momentum
$\vec{p}$ and angular
momentum $\vec{J}$ according to
$\vec{p}\to -\vec{p}$ and $\vec{J}\to \vec{J}$.
Consider Figure \ref{parity}.
\begin{stretchfigure}{1}{h}
\unitlength 0.01pt
\begin{picture}(40000,15000)(-5000,0)
\put(2000,0){\begin{picture}(15000,15000)
 \put(1000,7500){\vector(1,0){6200}}
 \put(4850,8100){\vector(-1,0){1000}}
 \put(-500,8700){${\bf V_1}(\vec{p_1},\la1)$}
 \put(14000,7500){\vector(-1,0){6200}}
 \put(10900,7800){\vector(0,1){1000}}
 \put(11000,6000){${\bf V_2}(\vec{p_2},\la2)$}
 \put(7700,7700){\vector(3,2){4600}}
 \put(9300,9500){\vector(3,2){1001}}
 \put(10000,11500){${\bf V_3}(\vec{p_3},\la3)$}
 \put(9300,7850){$\vartheta$}
 \put(7300,7300){\vector(-3,-2){4600}}
 \put(5800,5600){\vector(-3,-2){1001}}
 \put(-500,3000){${\bf V_4}(\vec{p_4},\la4)$}
\end{picture}}
\put(17500,0){\begin{picture}(5000,15000)
 \put(1000,7500){\vector(1,0){3000}}
 \put(2000,8500){P}
\end{picture}}
\put(24000,0){\begin{picture}(15000,15000)
 \put(1000,7500){\vector(1,0){6200}}
 \put(4100,7800){\vector(0,1){1000}}
 \put(-1000,9500){${\bf V_2}(\vec{p_1},-\la2)$}
 \put(14000,7500){\vector(-1,0){6200}}
 \put(11400,8100){\vector(-1,0){1000}}
 \put(11000,6000){${\bf V_1}(\vec{p_2},-\la1)$}
 \put(7700,7700){\vector(3,2){4600}}
 \put(10000,9900){\vector(-3,-2){1001}}
 \put(10000,11500){${\bf V_4}(\vec{p_3},-\la4)$}
 \put(9300,7850){$\vartheta$}
 \put(7300,7300){\vector(-3,-2){4600}}
 \put(5000,5000){\vector(3,2){1001}}
 \put(-500,3000){${\bf V_3}(\vec{p_4},-\la3)$}
\end{picture}}
\end{picture}\par
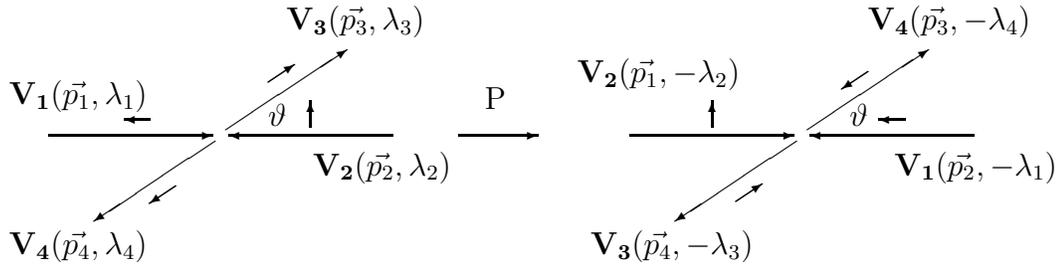
\caption{Transformation of initial and final states under a parity
 transformation \label{parity}}
\end{stretchfigure}
The small arrows symbolize the component
of spin in the direction of flight; from them,
the helicity can be read
off. For example, an arrow perpendicular to the direction of flight
designates a particle in the longitudinal polarization state.
Rotating the figure obtained after applying $P$ by an angle $\pi$
about the y-axis
one obtains the same physical situation as before
$P$ was applied, only
the particles' helicities have changed sign. It is thus clear that
$\M 1234(s,\vth)$ is equal to $\Mm 1234(s,\vth)$,
up to a possible phase factor.
The phase factor is found to be $(-1)^{\la3-\la4-\la1+\la2}$.
Rotating further by an angle $\pi$ about the z-axis
I obtain the relation
\be\M 1234(\vth)=\Mm1234(-\vth).\ee\par
When \underline{charge conjugation} is applied to
a state, particles are changed into
their antiparticles, while their momenta and helicities
remain unchanged.
The invariance of the $S$-matrix under charge conjugation
implies the relation
\be\M1234(AB\to CD)(\vth)=\M1234(\bar{A}\bar{B}
\to\bar{C}\bar{D})(\vth),\ee
where $\bar{A}$ is the antiparticle of particle $A$ etc.\par
After a rotation by an angle $\pi$ about the y-axis, an
exchange of particle labels in the
two-particle states and, succesively, a rotation by an angle $\pi$
about the z-axis, one
obtains the relation
\be\M1234(AB\to CD)(\vth)=\M2143(\bar{B}\bar{A}\to
\bar{D}\bar{C})(-\vth).\ee
\par
\begin{stretchfigure}{1}{ht}
\unitlength 0.01pt
\begin{picture}(40000,20000)(-5000,0)
\put(2000,7000){\begin{picture}(15000,15000)
 \put(1000,7500){\vector(1,0){6200}}
 \put(4100,7800){\vector(0,1){1000}}
 \put(1500,8700){\bf A}
 \put(14000,7500){\vector(-1,0){6200}}
 \put(11400,8100){\vector(-1,0){1000}}
 \put(13000,6000){\bf B}
 \put(7700,7700){\vector(3,2){4600}}
 \put(10000,9900){\vector(-3,-2){1001}}
 \put(12000,11500){\bf C}
 \put(9300,7850){$\vartheta$}
 \put(7300,7300){\vector(-3,-2){4600}}
 \put(5800,5600){\vector(-3,-2){1001}}
 \put(1500,3000){\bf D}
\end{picture}}
\put(17500,7000){\begin{picture}(5000,15000)
 \put(1000,7500){\vector(1,0){3000}}
 \put(2000,8500){\bf T}
\end{picture}}
\put(24000,7000){\begin{picture}(15000,15000)
 \put(7200,7500){\vector(-1,0){6200}}
 \put(4100,7800){\vector(0,1){1000}}
 \put(1500,8700){\bf A}
 \put(7800,7500){\vector(1,0){6200}}
 \put(10400,8100){\vector(1,0){1000}}
 \put(13000,6000){\bf B}
 \put(12360,10800){\vector(-3,-2){4600}}
 \put(9300,9500){\vector(3,2){1001}}
 \put(12000,11000){\bf C}
 \put(9300,7850){$\vartheta$}
 \put(2640,4200){\vector(3,2){4600}}
 \put(5000,5000){\vector(3,2){1001}}
 \put(1500,4000){\bf D}
\end{picture}}
\put(17500,-3000){\begin{picture}(5000,15000)
 \put(1000,7500){\vector(1,0){3000}}
 \put(2500,7500){\makebox(0,8000){Rotation by}}
 \put(2500,7500){\makebox(0,5000){angle $\pi-\vth$}}
 \put(2500,7500){\makebox(0,2000){about \^{y}}}
\end{picture}}
\put(24000,-3000){\begin{picture}(15000,15000)
 \put(1000,7500){\vector(1,0){6200}}
 \put(4350,8100){\vector(-1,0){1000}}
 \put(1500,8200){\bf C}
 \put(14000,7500){\vector(-1,0){6200}}
 \put(11400,8100){\vector(-1,0){1000}}
 \put(13000,6000){\bf D}
 \put(7300,7700){\vector(-3,2){4600}}
 \put(5700,9500){\vector(-3,2){1001}}
 \put(2000,11500){\bf B}
 \put(4900,7850){$\vartheta$}
 \put(7700,7300){\vector(3,-2){4600}}
 \put(9800,5600){\vector(-2,-3){800}}
 \put(10000,3500){\bf A}
\end{picture}}
\end{picture}
\caption[Transformation of initial and final states under
time reversal]
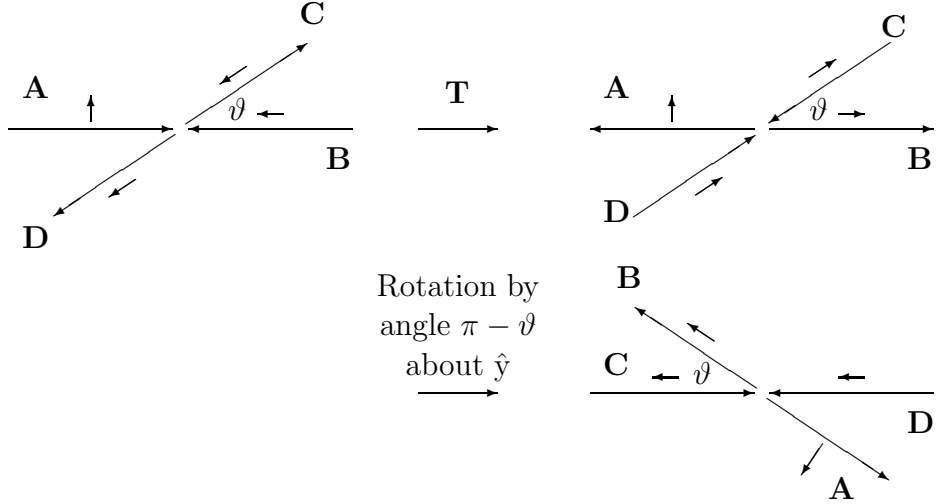
{Transformation of initial and final states under time reversal,
succeeded by a rotation \label{timerev}}
\end{stretchfigure}
\underline{Time reversal} changes initial to final states,
or, equivalently,
\[\left. \begin{array}{rcl}\vec{p}&\to&{-\vec{p}}\\
\vec{J}&\to&{-\vec{J}}
\end{array}\right\}\lambda\to\lambda,\]
where $\lambda$ denotes helicity (see Figure \ref{timerev}).
After a rotation
by an angle $\pi-\vth$ about the y-axis one obtains the relation
\be \M1234(AB\to CD)(\vth)=
\M3412(CD\to AB)({-\vth}).\ee\par
\subsubsection*{Application}
The ineractions of the GINDIS model are actually invariant under $P$-,
$C$- and $T$-transformations.\par
i)\quad\quad
Amplitudes for the processes
which I did not calculate -- listed in parentheses in \eref{massproc}
and \eref{photproc} -- can be
obtained from the ones which I calculated. For example, using
C-conjugation,
\[\M1234(W^+ W^+\to W^+ W^+)(\vth)=\M1234(W^- W^-\to W^- W^-)(\vth).\]
\par\noindent
ii)\quad\quad Amplitudes for a given process, $AB\to CD$, but
with different
helicities, can be related to each other
(cf. Table \ref{amprel}).
\begin{stretchtable}{1}{h}
\renewcommand{\arraystretch}{1.1}
\[\begin{array}{|c|c|r@{}l|}
\hline {\rm Condition\spc that}   & &\multicolumn{2}{|c|}{
{\rm Relation}}\\
{\rm is\spc fulfilled}&
\raisebox{1.5ex}[1.5ex]{\rm Transformation}&
\multicolumn{2}{|c|}{\M1234(\vth)=}\\
\hline ({\rm always})&P + {\rm rotations}&&\Mms1234({-\vth})\\
\hline \multicolumn{1}{|c|}{\,\quad A=\bar{B}}&&&\\
\multicolumn{1}{|c|}{\wedge\;C=\bar{D}}&
\raisebox{1.5ex}[1.5ex]{C + {\rm
rotations}}&&\raisebox{1.5ex}[1.5ex]{$\Ms2143({-\vth})$}\\ \hline
AB=CD&T + {\rm rotations}&&\Ms3412({-\vth})\\ \hline
{\rm Identical\spc initial}&{\rm Exchange\spc of}&&\\
{\rm partilces,\spc}A=B&{\rm labels +
rotation}&\raisebox{1.5ex}[1.5ex]{$(-1)^{\la1-\la2}$}
&\raisebox{1.5ex}[1.5ex]{$\Ms2134(\vth\pm\pi)$}\\
\hline{\rm Identical\spc final}&{\rm Exchange\spc of}&&\\
{\rm particles,\spc}C=D&{\rm labels +
rotation}&\raisebox{1.5ex}[1.5ex]{$(-1)^{\la4-\la3}$}&
\raisebox{1.5ex}[1.5ex]{$\Ms1243(\vth\pm\pi)$}\\
\hline
\end{array}\]
\caption[Relations among amplitudes]{Conditions, transformations
and relations among amplitudes for
a given process $AB\to CD$ using the Jacob and Wick phase
conventions.
The relations can be used to reduce the number of amplitudes to be
calculated for this process.\\
Sample usage:\\
For $W^-W^+\to W^-W^+$, we can use the invariance of
the $S$-matrix under
$P$-, $C$- and $T$-transformations.
One obtains relations due to $P$, $C$ and $T$ as well as
relations due to the
combined transformations
$CP$, $CT$, $PT$ and
$CPT$.\\
For example, due to $CPT$, ${\cal M}_{+--0}(\vth)={\cal
M}_{0++-}({-\vth})$, where the subscripts on ${\cal M}$ denote the
helicities.\label{amprel}}
\end{stretchtable}

Parity together with rotational invariance always gives a relation. In
addition, the fulfillment of each of the conditions
\begin{enumerate}
\item Particle A is the anti-particle of B ($A=\bar{B}$)
and particle C is the anti-particle of D ($ C=\bar{D}$).
\item The initial state contains the same particles as the
final state (in any order) ($AB=CD$).
\item\label{idin} The initial state contains identical
particles ($A=B$).
\item\label{idfin} The final state contains identical
particles ($C=D$).
\end{enumerate}
gives one more relation, each of which
follows from the invariance under a certain transformation, possibly
accompanied by rotations.
For the cases \ref{idin} and \ref{idfin} one obtains relations
among amplitudes in which identical particles have been exchanged.
It is
clear that the two amplitudes differ at most by a phase.
Finally,
all combinations of relations can also be applied to the
considered process.
\clearpage

\section{Listing of Amplitudes\label{amplist}}
\setcounter{figure}{0}
The amplitudes have been expanded in powers of $\tfrac{M_W^2}{s}\ll
1$. For the cases
\begin{enumerate}
\item $s\gg M_H^2$\label{sggmh}
\item no Higgs particle ($M_H\to\infty$)\label{nohiggs}
\end{enumerate}
I list the terms that grow as ${\cal O}(s)$. There are no terms that
grow with higher powers of $s$.
Case \ref{sggmh} is obtained by setting $\sti=\tti=\uti=0$.
For Case \ref{nohiggs} one has to set $\sti=-\tfrac{1}{4}$ and
$\tti=\uti=\tfrac{1}{2}$.
The terms depending on $\ew$ are in agreement with \cite{golare}.
The high-energy approximation has been carried out at a
fixed center-of-mass scattering angle $\vth$.
The expansion
breaks down in the
collinear region. More precisely, it is invalid if
$(1\pm\cos\vth)$ is so small
that it is comparable in magnitude to $\tfrac{M_{W,Z}^2}{s}$.
Terms bilinear in the $\ep{i}$ are not listed.
I omit these terms for consistency, because taking into
account bilinear terms
one would also have to consider terms of dimension
eight in the Lagrangian density,
since these are, like the bilinear terms, proportional
to $\Lambda^{-4}$.
\par
Amplitudes that are not listed are either related to one of the listed
amplitudes by one of the relations of Table \ref{amprel}
or do not have ${\cal O}(s)$-terms.
I note that, in Case \ref{sggmh}, no ${\cal O}(s)$-terms
are present whenever the standard
amplitude does not approach zero in the limit $s\to\infty$ except when
all external bosons are in the longitudinal polarization state.
In the listing,
\[\sg\equiv i\tfrac{g^2 s}{4\,M_W^2},\;\se\equiv
i\tfrac{e^2 s}{4\,M_W^2},\;
\sz\equiv i\tfrac{e^2 s}{4\,M_Z^2},\]
\[\swc{}\equiv\sw{},\quad\cwc{}\equiv\cw{},\quad\twc{}\equiv\tw{}.\]
\subsubsection*{$W^-W^+\to W^-W^+$}
\begin{eqnarray*} \Ml 0000 &=&
      -4\,(1-4\ewp)\sg\sti -(1-4\ewp)\sg\tti(1-\cos\vth)
      +3(\ewp+\twc2\ebp)\sg\,(1+\cos\vth)\\
\Ml 00++ &=&
     -8\ewp\,\sg\sti
     + \ew\,\sg\cos\vth - 2\ewp\,\sg\\
\Ml 0+0-&=&
      2\ewp\sg\tti(1-\cos\vth)
    + \left(\ewp-\tfrac{1}{2}\ew\right)\,\sg\cos\vth
    - (\ewp+\tfrac{3}{2}\ew)\,\sg\\
\Ml +++-&=&
    - 2\ew\,\sg(1+\cos\vth)\\
\Ml ++--&=&
     -4\ew\,\sg(1+\cos\vth)\\
\end{eqnarray*}
\subsubsection*{$W^+W^+\to W^+W^+$}
\begin{eqnarray*}
\Ml 0000&=&
   - (1-4\,\ewp)\sg\tti(1-\cos\vth)
   - (1-4\,\ewp)\sg\uti(1+\cos\vth)
   -6(\ewp+\twc2\ebp)\,\sg\\
\Ml 0+0-&=&
   2\ewp\,\sg\tti(1-\cos\vth)
   +\left(\ewp+\tfrac{1}{2}\ew\right)\,\sg\cos\vth
  -\left(\ewp-\tfrac{3}{2}\ew\right)\sg\\
\Ml +++-&=&
   4\ew\,\sg\\
\Ml ++--&=&
   8\ew\,\sg\\
\end{eqnarray*}
\subsubsection*{$W^-W^+\to Z\,Z$}
\begin{eqnarray*}
\Ml 0000&=&
 -4(1-4\ewp-2\twc2\ebp)\,\sg\sti
 + 6\ewp\,\sg\\
\Ml 00++&=&
   -8(\ewp+\twc2\ebp)\,\sg\sti
   +2(2\swc2(\ewp+\ebp)-\ewp-\twc2\ebp)\,\sg\\
\Ml 0+0-&=&
    -\tfrac{1}{2}(\ewp+\ebp)\,\sg\tfrac{\swc2}{\cwc{}}(1-\cos\vth)
 -\tfrac{1}{2}\ew\,\sg\cwc{}(3+\cos\vth)\\
\Ml ++00&=&
    -8\ewp\,\sg\sti
     -2\ewp\,\sg\\
\Ml +++-&=&\Ml +-++\;=\;
       -4\ew\,\sg\cwc2\\
\Ml ++--&=&
       -8\ew\,\sg\cwc2\\
\end{eqnarray*}
\subsubsection*{$W^-Z\to W^-Z$}
\begin{eqnarray*}
\Ml 0000&=&
   -(1+4\ewp-2\twc2\ebp)\,\sg\tti(1-\cos\vth)
 -3 \ewp\,\sg(1-\cos\vth)\\
\Ml 00++&=&
 - (\ewp+\ebp)\,\sg\tfrac{\swc2}{\cwc{}}
  + \ew\,\sg\cwc{}\cos\vth\\
\Ml 0+0-&=&
2(\ewp+\twc2\ebp)\sg\tti(1-\cos\vth)
-2(\ewp+\ebp)\,\sg\cwc2(1-\cos\vth)\\
&&-\ebp\,\sg\tfrac{1-\cos\vth}{\cwc2}
  +(\ewp+3\ebp)\,\sg  (1-\cos\vth)\\
\Ml 0+-0&=&
\tfrac{1}{2} (\ewp+\ebp)\,\sg\tfrac{\swc2}{\cwc{}}(1+\cos\vth)
-\tfrac{1}{2} \ew\,\sg\cwc{}(3-\cos\vth)\\
\Ml +0-0&=&
2\ewp\,\sg\tti(1-\cos\vth)
-\ewp\,\sg(1-\cos\vth)\\
\Ml +++-&=&\Ml ++-+\;=\;
2\ew\,\sg\cwc2(1-\cos\vth)\\
\Ml ++--&=&
4\ew\,\sg\cwc2  ( 1-\cos\vth )\\
\end{eqnarray*}
\subsubsection*{$Z\,Z\to Z\,Z$}
\begin{eqnarray*}
\Ml 0000&=&
(1-4\cwc2\ewp-4\swc2\ebp)\sg\left(-4\sti -\tti(1-\cos\vth)
-\uti(1+\cos\vth)\right)\\
\Ml 00++&=&
   -8(\ewp+\twc2\ebp)\,\sg\sti
     -2(\ewp+\twc2\ebp)\,\sg\\
\Ml 0+0-&=&
 2(\ewp+\twc2\ebp)\,\sg\tti(1-\cos\vth)
  -(\ewp+\twc2\ebp)\,\sg(1-\cos\vth)\\
\end{eqnarray*}
\subsubsection*{$\gamma\,\gamma\to W^-W^+$}
\begin{eqnarray*}
\Ml ++00&=&
     -4(\ewp+\ebp)\,\se\\
\Ml +++-&=&\Ml +-++\;=\;
       -4\ew\,\se\\
\Ml ++--&=&
     -8\ew\,\se\\
\end{eqnarray*}
\subsubsection*{$\gamma\,W^-\to\gamma\,W^-$}
\begin{eqnarray*}
\Ml +0-0&=&
    -2(\ewp+\ebp)\,\se(1-\cos\vth)\\
\Ml +++-&=&\Ml ++-+\;=\;
      2\ew\,\se (1-\cos\vth)\\
\Ml ++--&=&
      4\ew\,\se(1-\cos\vth)\\
\end{eqnarray*}
\subsubsection*{$\gamma\,Z\to W^- W^+$}
\begin{eqnarray*}
\Ml +00-&=&
    -\tfrac{1}{2}(\ewp+\ebp)\tfrac{\se}{\swc{}}(1+\cos\vth)
    -\tfrac{1}{2}\ew\tfrac{\se}{\swc{}}(-3+\cos\vth)\\
\Ml +0-0(\vth)&=&
     -\Ml +00-(\pi{-}\vth)\\
\Ml ++00&=&
     -4\,(\ewp-\ebp)\tfrac{\se\sti}{\swc{}\cwc{}}
     +(\ewp-\ebp)\tfrac{\se}{\swc{}\cwc{}}
     -4\,(\ewp-\tw2\ebp)\,\se\tfrac{\cwc{}}{\swc{}}\\
\Ml +++-&=&\Ml ++-+\;=\;\Ml +-++\;=\;\Ml +---\;=\;
     -4\,\ew\,\se\tfrac{\cwc{}}{\swc{}}\\
\Ml ++--&=& -8\,\ew\,\se\tfrac{\cwc{}}{\swc{}}\\
\end{eqnarray*}

\end{appendix}


\begin{thebibliography}{99}
\addcontentsline{toc}{chapter}{Bibliography}
\bibitem{sm}S. L. Glashow, Nucl. Phys. {\bf 22} (1961) 579;
S. Weinberg, Phys. Rev. Lett. {\bf 19} (1967) 1264;
A. Salam, Proc. 8th Nobel Symposium, ed. N. Svartholm
(Almquits and Wiksells,
Stockholm, 1968), p.~367
\bibitem{thooft}G. 't Hooft, \nphb{35} (1971) 167; B. W. Lee, \phrd{5}
(1972) 823
\bibitem{llewellyn}C. H. Llewellyn Smith, \phlb{46} (1973) 233
\bibitem{joglekar}S. D. Joglekar, Annals of Physics {\bf 83}
(1974) 427
\bibitem{clt}J. M. Cornwall, D. N. Levin and G. Tiktopoulos, \phrd{10}
(1974) 1145
\bibitem{velt}M. Veltman, Acta Phys. Polon. {\bf B12} (1981) 437
\bibitem{kmss}J. Maalampi, D. Schildknecht and K. H. Schwarzer,
\phlb{166} (1986) 361;
M. Kuroda, J. Maalampi, D. Schildknecht and K. H. Schwarzer,
Nucl. Phys. {\bf B284} (1987) 271;
M. Kuroda, J. Maalampi, D. Schildknecht and K. H. Schwarzer,
\phlb{190} (1987) 217
\bibitem{bks}C. Bilchak, M. Kuroda and D. Schildknecht, Nucl. Phys.
{\bf B299} (1988) 7
\bibitem{nogauge}E. Yehudai, \phrd{44} (1991) 3434;
G. Gounaris, J. L. Kneur, J. Layssac, G. Moultaka, F. M. Renard
and D. Schildknecht (p.735) and
G. B\'{e}langer and F. Boudjema (p.763), in
``$e^+e^-$ Collisions at 500 GeV, The Physics Potential'',
ed.\ P. W. Zerwas, DESY-Preprint DESY 92-123 (1992);
G. B\'{e}langer and F. Boudjema, \phlb{288} (1992) 201;
C. Grosse-Knetter and D. Schildknecht, \phlb{302} (1993) 309;
O. J. P. \'{E}boli, M. C. Gonzal\'{e}z-Garc\'{i}a and S. F. Novaes,
Madison-Preprint MAD/PH/764 (1993), hep-ph/9306306
\bibitem{gg}K. Gaemers and G. Gounaris, Z. Phys. {\bf C1} (1979) 259;
K. Hagiwara, R. D. Peccei, D. Zeppenfeld and K. Hisaka,
Nucl. Phys. {\bf B282} (1987) 253
\bibitem{bkrs}M. Bilenky, J. L. Kneur, F. M. Renard and
D. Schildknecht,
\nphb{409} (1993) 22, and literature cited there
\bibitem{chiral}B. Holdom, \phlb{258} (1991) 156; A. Falk, M. Luke and
E. Simmons, \nphb{365} (1991) 523; D. Espriu and M. Herrero,
\nphb{373}
(1991) 117;
J. Bagger, S. Dawson and G. Valencia, \nphb{399} (1993) 364
\bibitem{cvekoe}G. Cveti\v{c}, R. K\"{o}gerler, \nphb{363} (1991) 401
\bibitem{gkko}C. Grosse-Knetter and R. K\"ogerler, \phrd{48} (1993)
2865
\bibitem{buwyllr}C. N. Leung, S. T. Love and S. Rao,
\zphc{31} (1986) 433;
W. Buchm\"uller and D. Wyler, \nphb{268} (1986) 621
\bibitem{quad}M. Kuroda, F. M. Renard and
D. Schildknecht, \phlb{183} (1987)
366
\bibitem{gore}G. J. Gounaris and F. M. Renard, \zphc{59} (1993) 133
\bibitem{ruj}A. de R\'{u}jula, M. B. Gavela, P.
Hern\'{a}ndez and E. Mass\'{o},
\nphb{384} (1992) 3
\bibitem{haze}K. Hagiwara, S. Ishihara, R. Szalapski
and D. Zeppenfeld,
\phlb{283} (1992) 353; \phrd{48} (1993) 2182
\bibitem{heve}P. Hern\'{a}ndez and F. J. Vegas, \phlb{307} (1993) 116
\bibitem{eiwud}M. B. Einhorn and J. Wudka,
Michigan Preprint UM-TH-92-25
(1992);
J. Wudka, UC-Riverside Preprint UCRHEP-108 (1993),
hep-ph/9305292
\bibitem{anom}C. Grosse-Knetter, I. Kuss and D. Schildknecht,
\zphc{60} (1993) 375
\bibitem{jacobwick}M. Jacob and G. C. Wick, Annals of Physics {\bf 7}
(1959) 404
\bibitem{dipl}I. Kuss (Diploma Thesis),
Bielefeld-Preprint BI-TP 93/57 (1993)
\bibitem{appe}T. Appelquist and C. Bernard, \phrd{22} (1980) 200;
A. C. Longhitano, \nphb{188} (1981) 118; R. Casalbuoni, S. de Curtis,
D. Dominici and R. Gatto, \nphb{282} (1987) 235
\bibitem{santa}M. Bilenky and A. Santamaria,
CERN-Preprint CERNT-TH.7030/93
and Bielefeld-Preprint BI-TP 93/48
\bibitem{kmsscut}H. Neufeld, J. D. Stroughair and D. Schildknecht,
\phlb{198} (1987) 563
\bibitem{grs}G. Gounaris, F. M. Renard, D. Schildknecht, Phys. Lett.
{\bf B263} (1991) 291
\bibitem{symmetries}J. J. Sakurai, Invariance Principles
and Elementary
Particles, Princeton University Press (1964);
J. Werle, Relativistic Theory of Reactions,
Polish Scientific Publishers
(1966)
\bibitem{golare}G. J. Gounaris, J. Layssac and F. M. Renard,
Montpellier-Preprint PM/93-26 and Thessaloniki-Preprint THES-TP 93/8
\end{thebibliography}
\end{document}